\documentclass[aps,prl,twocolumn,showpacs,preprintnumbers,amsmath,amssymb,floatfix,nofootinbib]{revtex4-1}
\usepackage{graphicx}
\usepackage{color}
\usepackage{colordvi}

\usepackage[normalem]{ulem}

\usepackage{color}
\newcommand{\beqn}{\begin{eqnarray}}
\newcommand{\eeqn}{\end{eqnarray}}
\newcommand{\be}{\begin{equation}}
\newcommand{\ee}{\end{equation}}

\newcommand{\ba}{\begin{array}{c}}
\newcommand{\bat}{\begin{array}{cc}}
\newcommand{\ea}{\end{array}}

\newcommand{\bi}{\begin{itemize}}
\newcommand{\ei}{\end{itemize}}

\newcommand{\ket}{\,\rangle}
\newcommand{\bra}{\langle \,}

\newcommand{\Frac}[2]{\frac{\displaystyle #1}{\displaystyle #2}}

\newcommand{\cO}{{\cal O}}

\newcommand{\mF}{\mathcal{F}}

\newcommand{\mK}{\mathcal{K}}
\newcommand{\mL}{\mathcal{L}}
\newcommand{\mM}{\mathcal{M}}
\newcommand{\mO}{\mathcal{O}}

\newcommand{\Int}{\displaystyle{\int}}

\newcommand{\bear}{\begin{eqnarray}}
\newcommand{\eear}{\end{eqnarray}}

\newcommand{\divergence}{\lambda}
\newcommand{\nn}{\nonumber}

\newcommand{\sla}{\hspace*{-0.2cm}\slash  }

\begin{document}

\preprint{FTUAM-15-16}
\preprint{IFT-UAM/CSIC-15-058}
\preprint{TUM-HEP-997/15}


\title{One-loop renormalization of the electroweak chiral Lagrangian with a light Higgs}

\author{Feng-Kun Guo $^{1}$}

\author{Pedro Ruiz-Femen\'\i a  $^{2}$}

\author{Juan Jos\'e Sanz-Cillero $^{3}$}

\affiliation{${}^1$ Helmholtz-Institut f\"ur Strahlen- und Kernphysik and Bethe Center
for Theoretical Physics, Universit\"at Bonn, D-53115 Bonn, Germany }
\affiliation{${}^2$
  Physik Department T31, James-Franck-Strasse, Technische Universit\"at M\"unchen, D-85748 Garching, Germany         }
\affiliation{${}^3$
Departamento de F\'\i sica Te\'orica and Instituto de F\'\i sica Te\'orica, IFT-UAM/CSIC,
Universidad Aut\'onoma de Madrid, Cantoblanco, 28049 Madrid, Spain }

\begin{abstract}
We consider the general chiral effective action which parametrizes the
nonlinear realization of the
spontaneous breaking of the electroweak symmetry with a light Higgs, and
compute the one-loop ultraviolet divergences coming from Higgs and
electroweak
Goldstone fluctuations using the background field method. The renormalization of the
divergences is carried out through operators of
next-to-leading order in the chiral counting, i.e. of ${\cal O}(p^4)$.
Being of the same order in power counting, the
logarithmic corrections linked to these divergences
can be as important as the tree-level contributions from
the ${\cal O}(p^4)$ operators, and must be accounted for in the phenomenological
analysis of experimental data.
Deviations  in the  ${\cal O}(p^2)$ (leading-order)
couplings with respect to the Standard Model values, {\it e.g.}, in
the $h\to WW$ coupling, would generate contributions from the
1-loop chiral logarithms computed in this work
 to a vast variety of observables,
which do not have a counterpart in the conventional electroweak effective theory with a
linearly transforming Higgs complex doublet.
\end{abstract}

\maketitle

\section{Introduction}

The lack of experimental evidence of the new physics states predicted by
many  natural solutions to the electroweak (EW) symmetry breaking has brought an
increasing interest in the analysis of possible deviations from the Standard
Model (SM) using broader frameworks.
Assuming there is a large energy gap between the EW scale $v = 246$~GeV and the new physics scale,
a description of the EW symmetry breaking (EWSB) in general terms is
provided by an effective field theory (EFT) built from the presently known
particle content, including a light Higgs scalar, and based on
the spontaneous symmetry breaking pattern
of the SM: $G=SU(2)_L\times SU(2)_R$ breaks down to the custodial group
$H=SU(2)_{L+R}$,  where $SU(2)_L\times U(1)_Y\subset G$ is gauged
and the three would-be Goldstone bosons $\pi^a$ that arise from the spontaneous EWSB
give mass to the $W^\pm,Z$ in the unitary gauge.

The existence of an approximate custodial symmetry guarantees that the
$\rho$-parameter corrections are small, the latter being originated from
radiative corrections or (custodial-symmetry breaking) operators that are
subleading in the EFT power-counting~\cite{custodial}.
Without loss of generality, the Higgs $h$ will be taken to be a singlet under
the full group $G$ and the Goldstones are nonlinearly realized in our EFT
approach~\cite{CCWZ};
linear models with a Higgs complex doublet $\Phi$ (e.g. the SM) are just a
subset within the general class of nonlinear theories, as it is always possible
to express $\Phi$ in terms of the singlet $h$ and the nonlinearly realized
Goldstones.
Indeed, there is a vast variety of beyond-SM theories where the Higgs is a
composite particle, typically a pseudo-Goldstone of some type, and shows the
characteristic nonlinear interaction structure of this kind of particles (see
the review~\cite{composite-rev}).

The framework described above shares many similarities with the low-energy limit
of QCD, also ruled by the $SU(2)_L\times SU(2)_R$ chiral symmetry, and we can
expect that well-known aspects from QCD are reproduced in the nonlinear EW
chiral EFT. In particular, we are interested in this work
in
the relevant role of the EW chiral logarithms arising from radiative corrections.
We know from their QCD analogues that such contributions to
one-loop amplitudes are in many cases as important as the tree-level
contributions from higher dimension operators, due to the nonlinear structure of
the Goldstone interactions (that is the case for instance in $\pi\pi$ scattering
in the scalar-isoscalar channel~\cite{chpt}).

Motivated by this fact, in this letter we study the radiative corrections coming
from scalar boson loops (Higgs and EW Goldstones) within the framework of a
nonlinear EW chiral Lagrangian including a light Higgs (ECLh).
Ref.~\cite{1loop-linear} computed the ultraviolet (UV) divergence at
next-to-leading order (NLO) in the linear Higgs EFT. The present work complements that analysis
and provides the UV divergences and the renormalization for the nonlinear case
at NLO in the low-energy chiral counting, i.e. at $\cO(p^4)$,
where $p$ denotes low-energy scales, either light-particle masses or momenta.
The ECLh contains all the SM particles and it is expected to
describe their interaction at energies much below the cut-off of the EFT,
$\Lambda_{\rm ECLh}$, given by either the scale at which one encounters a new
state or the energy where loop corrections turn too large to validate a
perturbative expansion, naively $4\pi v\approx 3$~TeV.

Based on dimensional analysis it is possible to organize a low-energy
expansion of the amplitudes in powers of low-energy scales $p$~\cite{Weinberg:1979,Georgi-Manohar}
and implement a chiral power counting in the nonlinear EFT
Lagrangian~\cite{Hirn:2004,EW-chiral-counting,1loop-AA-scat}.
Renormalization is then carried out order-by-order in the low-energy
expansion~\cite{chpt,Morales:94,Longhitano:1980iz,1loop-Gavela}.

The one-loop effective action  shows a series of UV divergences of higher
dimension that require the inclusion of new operators in the Lagrangian, which
are NLO in the chiral counting.
While the precise value of the NLO couplings depends on the underlying physics,
their running is fully determined by the leading order (LO) Lagrangian and its
symmetry structure.
Theoretical predictions that incorporate the latter operators must therefore
account also for the logarithmic contributions in order to reach the percent
accuracy level in the data analyses of next runs at present and future
colliders~\cite{collider-prospects}. Unless assumptions about the physics
underlying the ECLh are taken, such phenomenological studies must  account in
full generality for the tree-level contributions from NLO
operators~\cite{EW-chiral-counting,ECLh-Gavela}, the running and the associated
NLO logarithms (that are provided in this letter), and the NLO loop finite
pieces~\cite{gg-ZZ,gg-hh,Azatov:2013,1loop-WW-scat,1loop-AA-scat}, which vary
from one observable to another.

\section{Low-energy effective theory}

At low energies the amplitudes can be then organized in terms of
a chiral expansion in powers of the low-energy scales $p$.
Analyzing the contributions from the effective operators to the
amplitudes it is possible to establish a consistent (chiral) power
counting for the ECLh Lagrangian~\cite{Hirn:2004,EW-chiral-counting,1loop-AA-scat,Hirn:2004},
which is organized as
\bear
\mL &=& \mL_2 \, + \, \mL_4\, +  ...\, ,
\label{eq.EFT-Lagr}
\eear
where $\mL_n$ contains terms that scale as $\cO(p^n)$.
The LO Lagrangian is given
by~\cite{EW-chiral-counting,ECLh-Gavela,ECLh-other,SILH},
\bear
\mL_2 &=& \Frac{v^2}{4} \mF_C \bra u_\mu u^\mu \ket
+\Frac{1}{2} (\partial_\mu h)^2 -
v^2 \, V
\nn\\
&& + \mL_{YM} +   i \bar{\psi}\,   D \sla \, \psi \, -\, v^2 \bra  J_{S}\ket \, ,
\label{eq.L2}
\eear
where $\bra ...\ket$
stands for the trace of the $2\times 2$ EW tensors, $\mL_{YM}$ is the
Yang--Mills Lagrangian for the gauge fields,
$D$ is the gauge covariant derivative acting on the fermions, and $J_S$
denotes the Yukawa coupling of the fermions to the Higgs and Goldstone fields
defined below.
The factors of $v$ in the normalization of some terms are  introduced for later convenience.
$\mF_C,\, V$ and $J_S$ are functionals of $x=h/v$, and
have Taylor expansions
\bear
&&  \mF_C[x]  = 1 + 2 a x + b x^2 +...\, ,
\qquad
 J_S[x]=\sum_n J_S^{(n)} x^n/n! \, ,
\nn\\
&&
V[x]= m_h^2 \left( \frac{1}{2}  x^2+\Frac{1}{2} d_3 x^3 + \Frac{1}{8} d_4 x^4 \, +  ... \right)\, ,
\eear
given in terms of the constants
$a,b,m_h$, etc.~\cite{EW-chiral-counting,ECLh-Gavela,ECLh-other,SILH},
with $J_S^{(n)}$
the $n$-th derivative with respect to $h/v$.
In the nonlinear realization of the spontaneous EWSB
the Goldstones are parameterized through the coordinates  $(u_L,u_R)$ of the
$SU(2)_L\times SU(2)_R/SU(2)_{L+R}$ coset space~\cite{CCWZ}, with the unitary matrices $u_{L,R}$
being functions of the Goldstone fields $\pi^a$  which enter through the building blocks
\bear
u^{}_\mu &=&
i u_R^\dagger (\partial^{}_\mu-i r^{}_\mu) u^{}_R - iu_L^\dagger (\partial^{}_\mu-i \ell^{}_\mu) u^{}_L\, ,
\nn\\
\Gamma^{}_\mu &=& \frac{1}{2} u_R^\dagger (\partial^{}_\mu-i r^{}_\mu) u^{}_R
+\frac{1}{2}u_L^\dagger (\partial^{}_\mu-i \ell^{}_\mu) u^{}_L\, ,
\nn\\
f_\pm^{\mu\nu}
&=& u_L^\dagger \ell^{\mu\nu} u_L \pm u_R^\dagger r^{\mu\nu} u^{}_R\, ,
\eear
where $r_{\mu\nu} = \partial_\mu r_\nu - \partial_\nu r_\mu -i [r_\mu ,r_\nu]$,
and the left-hand counter part $\ell_{\mu\nu}$ is defined analogously.
The  tensor $J_S$ is defined as
\bear
  J^{}_{S} &=& J^{}_{YRL} +J_{YRL}^\dag, \nn\\
  J^{}_{YRL} &=& - \Frac{1}{\sqrt{2} v} u_R^\dagger \,   \, \hat{Y}\,   \,
\psi^{\alpha}_R \bar{\psi}^{\alpha}_L
u^{}_L  \, ,
\eear
where $\psi_{R,L}=\frac12(1\pm \gamma^5)\psi$ and $\psi = (t,b)^T$ is the
top-bottom SM $SU(2)$ doublet.
The inclusion of additional doublets is straightforward and will be discussed later.
The summation over the Dirac index $\alpha$ in $\psi^{\alpha}_R \bar{\psi}^{\alpha}_L$
is assumed and its tensor structure under $G$ is left implicit. The $2\times 2$ matrix
$\hat{Y}[h/v]$ is a spurion auxiliary field, functional of $h/v$,
which incorporates the fermionic Yukawa coupling, allowing the inclusion
of explicit custodial symmetry-breaking terms~\cite{Hirn:2004}.

The low-energy chiral counting of the building blocks is provided by the
scaling $\{\partial_\mu, r_\mu, \ell_\mu, m_{h,W,Z,\psi}\}\sim \cO(p)$,
$\{\pi^a/v,\,    u_{L,R},\,       h/v \}\sim \cO(p^0)$,
and $\psi/v\sim O(p^{1/2})$~\cite{Georgi-Manohar,Weinberg:1979}.  Accordingly,
covariant derivatives must scale as the ordinary ones and $g,g',\hat{Y}\sim \cO(p/v)$.
Since this implies $\mF_C\sim \cO(p^0)$,
$u_\mu \sim \cO(p)$ and  $f_\pm^{\mu\nu},\, J_{YRL},\, J_S,\, V \sim \cO(p^2)$,
the LO Lagrangian in Eq.~(\ref{eq.L2}) is $\cO(p^2)$ and the
one-loop corrections are formally
$\cO(p^4)$~\cite{1loop-AA-scat,Hirn:2004,EW-chiral-counting,Georgi-Manohar,Weinberg:1979}.

The transformations of the building blocks under $G$ are given by
\bear
h &\rightarrow & h \, , \qquad
Y \rightarrow  g^{}_R Y g_R^\dagger \, , \qquad
u^{}_{R/L} \rightarrow  g^{}_{R/L} \, u^{}_{R/L} \, g_h^\dagger \, ,\nn
\\
r^{}_\mu &\rightarrow & g^{}_R r_\mu g_R^\dagger + i g^{}_R \partial^{}_\mu
g_R^\dagger \, , \qquad
\ell^{}_\mu \rightarrow  g^{}_L \ell_\mu g_L^\dagger + i g^{}_L \partial^{}_\mu
g_L^\dagger \, ,
\nn\\
\mO
&\rightarrow &
g^{}_h \, \mO \,  g_h^\dagger\, ~~\text{for}~~\mO\in \{ \, u^{}_\mu
 , \, J_{YRL}
, \, J^{}_{S}\, , \, 
f_\pm^{\mu\nu}\,
\}\, ,
\label{eq.transformations}
\eear
where $g_{R/L}\in SU(2)_{R/L}$, and $g_h\in SU(2)_{L+R}$ is the compensating
transformation of the Goldstone representatives in the CCWZ
formalism~\cite{CCWZ}.

The SM is recovered by setting $\mF_C=(1+h/v)^2$,
$V= \frac{1}{4} \lambda v^2 [(1+h/v)^2-1]^2$
and $\hat{Y}= (1+h/v) \, (y_t P_+ + y_b P_-)$,
defined in terms of the
SM Yukawa coupling constants $y_{q}$
and the projectors $P_\pm=(1\pm\sigma^3)/2$
($\sigma^a$ are the Pauli matrices).
Other SM fermion doublets and the
flavor symmetry breaking between generations can be incorporated
by adding in $J_{YRL}$ an additional family index in the fermion fields,
$\psi^A$, and promoting $\hat{Y}$ to a tensor $\hat{Y}^{AB}$
in the generation space~\cite{flavor-ECLh}.

In our analysis, $\ell_\mu,\, r_\mu , \, \hat{Y}$ are spurion
auxiliary background fields
that keep the invariance of the ECLh action under $G$.
When evaluating physical matrix elements, custodial symmetry is then broken in the same way as in the SM,
keeping only the gauge invariance under the subgroup
$SU(2)_L\times U(1)_Y\subset G$~\cite{Longhitano:1980iz,ECLh-Gavela,ECLh-other,EW-chiral-counting}:
\begin{gather}
\ell_\mu = - \Frac{g}{2} W_\mu^a \sigma^a\, , \quad
r_\mu = - \Frac{g'}{2} B_\mu \sigma^3\, , \quad
\nn\\
\hat{Y} =  \hat{y}_t P_+\, +\, \hat{y}_b P_- \, ,
\end{gather}
where $\hat{y}_{t,b}$ must be also understood as functionals of $h/v$.
Notice that the Yukawa operators are still invariant under global $G$ transformations
if $\hat{y}_t=\hat{y}_b$.

\section{Effective action at one loop}

Our aim is to compute the one-loop UV divergences of the effective action
by means of the background field method~\cite{heat-kernel}.
We choose for the $G/H$ coset representatives $u_L=u_R^\dagger$~\cite{Pich:2013}
and perform fluctuations of the scalar fields (Higgs and Goldstones)
around the classical background fields  $\bar{h}$ and
$\bar{u}_{L,R}$, respectively, in a way analogous to Ref.~\cite{rcht-integral}:
\bear
u_{R,L}&=& \bar{u}_{R,L} \, \exp\{ \pm i \mF_C^{-1/2}\,     \Delta/(2 v)     \}\, ,
\,\,\,\,\,
h = \bar{h}\, +\, \epsilon\, ,
\eear
with $\Delta=\Delta^a \sigma^a$. Without any loss of generality we have introduced the factor $\mF_C^{-1/2}$
in the exponent for later convenience; it will allow us to write down the second-order fluctuation of the action
in the {canonical form}~\cite{heat-kernel}.

To obtain the one-loop effective action within the background field method
one retains  the quantum fluctuations $\vec{\eta}^T=(\Delta^a, \epsilon)$
up to quadratic order~\cite{heat-kernel},
\bear
\mL_2 &=& \mL_2^{\cO(\eta^0)}  \, +\,  \mL_2^{\cO(\eta^1)} \, +\,  \mL_2^{\cO(\eta^2)}  \, +\,  \cO(\eta^3)\, ,
\label{eq.Lagr-fluctuation}
\eear
where $\mL_2^{\cO(\eta^0)}=\mL_2[\bar{u}_{L,R},\bar{h}]$.
The tree-level effective action is equal to the  action evaluated at the classical solution,
$\int {\rm d^dx} \, \mL^{\cO(\eta^0)}$, where $\int {\rm d^dx} \, \mL_2^{\cO(\eta^0)}$
denotes the LO contribution in the chiral expansion.
The background field configurations $\bar{h}$ and $\bar{u}_{L,R}$
correspond to the solutions of the classical equations of motion (EoM),
defined by the vanishing of the linear term $\mL_2^{\cO(\eta)}$ for
arbitrary $\vec{\eta}$. They read
\bear
\nabla^\mu u_\mu &=& -2 J_{P} / \mF_C  - u_\mu \partial^\mu
(\ln\mF_C)
\, ,
\nn\\
\partial^2 h/v &=& \Frac{1}{4} \mF_C'   \, \bra u_\mu u^\mu \ket
   - V'
 -\,\bra  J_{S}'\ket \, ,
\label{eq.EoM}
\eear
with $J^{}_{P} = i(J^{}_{YRL} -J_{YRL}^\dag)$ and
the covariant derivative  $\nabla_\mu \cdot =\partial_\mu + [\Gamma_\mu ,\cdot]$.
Here and  in the following, we abuse of the notation by writing the
background fields $\bar{u}_{\mu}$ and $\bar{h}$ as ${u}_{\mu}$ and ${h}$ for
conciseness.
 
\begin{table*}[!t]
\begin{center}
\renewcommand{\arraystretch}{1.5}
\begin{tabular}{|c|c|c|c|}
\hline
 $c_k$  &
Operator ${\cal O}_k$ &
$\Gamma_k$ & $\Gamma_{k,0}$  \\
\hline
\hline
  $c_1$   &  
$\Frac{1}{4}\bra {f}_+^{\mu\nu} {f}_{+\, \mu\nu}
- {f}_-^{\mu\nu} {f}_{-\, \mu\nu}\ket$
&  $\Frac{1}{24}(\mK^2-4)$
& $-\Frac{1}{6}(1-a^2)$
\\
\hline
   $(c_2 -c_3)$    &  
$\frac{i}{2}  \bra {f}_+^{\mu\nu} [u_\mu, u_\nu] \ket$
&  $\Frac{1}{24}(\mK^2-4)$
& $-\Frac{1}{6}(1-a^2)$
\\
\hline
  $c_4$   &   
$\bra u_\mu u_\nu\ket \, \bra u^\mu u^\nu\ket $
&  $\Frac{1}{96}(\mK^2-4)^2$
& $\Frac{1}{6}(1-a^2)^2$
\\
\hline
   $c_5$     &    
$ \bra u_\mu u^\mu\ket^2$
&  $\Frac{1}{192} (\mK^2-4)^2 + \Frac{1}{128} \mF_C^2 \Omega^2$
& $\Frac{1}{8}(a^2-b)^2 + \Frac{1}{12} (1-a^2)^2$
    \\
\hline
  $c_6$    &  
$\Frac{1}{v^2}(\partial_\mu h)(\partial^\mu h)\,\bra u_\nu u^\nu \ket$
&  $\Frac{1}{16}\Omega  (\mK^2-4) - \Frac{1}{96} \mF_C \Omega^2  $
& $-\Frac{1}{6}(a^2-b)(7a^2 -b-6)$
     \\
\hline
$c_7$     &   
$\Frac{1}{v^2}(\partial_\mu h)(\partial_\nu h) \,\bra u^\mu u^\nu \ket$
&  $\Frac{1}{24}\mF_C \Omega^2 $
& $\Frac{2}{3}(a^2-b)^2$
    \\
\hline
$c_8$      &  
$\Frac{1}{v^4}(\partial_\mu h)(\partial^\mu h)(\partial_\nu h)(\partial^\nu h)$
&  $\Frac{3}{32}\Omega^2$
& $\Frac{3}{2}(a^2-b)^2$
     \\
\hline
$c_9$    &  
   $\Frac{(\partial_\mu h)}{v}\,\bra f_-^{\mu\nu}u_\nu \ket$
&  $\Frac{1}{24} \mF_C' \Omega$
&  $-\Frac{1}{3}a (a^2-b)$
\\
\hline
  $c_{10}$    & 
$  \Frac{1}{2} \bra {f}_+^{\mu\nu} {f}_{+\, \mu\nu}
+ {f}_-^{\mu\nu} {f}_{-\, \mu\nu}\ket$
&  $-\Frac{1}{48}(\mK^2+4)$
& $-\Frac{1}{12}(1+a^2)$
\\
\hline
\end{tabular}
\end{center}
\caption{{\small
Purely bosonic operators needed for the renormalization of
the NLO effective Lagrangian $\mL_4$. In the last column, we provide
the first term $\Gamma_{k,0}$ in the expansion of the $\Gamma_k$ in powers of
$(h/v)$ by using $\mF_C=1+2 a h/v + b h^2/v^2 +\cO(h^3)$
(no further term is needed).
The first five operators $\mO_{i}$
have the structure of the respective $a_i$ Longhitano operator~\cite{Longhitano:1980iz,Morales:94}
(with $i=1...5$). In addition, $c_6=\mF_{D7}$, $c_7=\mF_{D8}$  and $c_8=\mF_{D11}$ in the notation
of Ref.~\cite{EW-chiral-counting}. The last operator of the list,
$\mO_{10}=2\bra r_{\mu\nu} r^{\mu\nu}+\ell_{\mu\nu}\ell^{\mu\nu}\ket$, only
depends on the EW field strength tensors  and its coefficient is labeled as
$c_{10}=H_1$ in the notation of Ref.~\cite{chpt}.
}}
\label{tab.div}
\end{table*}

The quadratic fluctuation $\mL_2^{\cO(\eta^2)}$ reads
\bear
\mL^{\cO(\Delta^2)} &=&
-\, \Frac{1}{4} \bra \Delta \nabla^2 \Delta\ket
\,  + \, \Frac{1}{16}\bra  [u_\mu,\Delta] \, [u^\mu, \Delta] \ket
\nn\\
&&
\hspace*{-1.7cm}+  \left[  \Frac{  \mF_C^{-\frac{1}{2}} \mK}{8}
\left(\Frac{\partial^2 h}{v}\right)
+ \Frac{\Omega}{16} \left(\Frac{\partial_\mu h}{v}\right)^2 \right]
   \bra \Delta^2 \ket
+ \Frac{1}{2\mF_C} \bra \Delta^2 J_S\ket
 ,
\nn\\
\mL^{\cO(\epsilon^2)} &= &
-\Frac{1}{2} \epsilon \left[    \partial^2
 - \Frac{1}{4} \mF_C''    \bra u_\mu u^\mu \ket
+     V''
+ \bra J_S''\ket
\right]\, \epsilon\, ,
\nn\\
\mL^{\cO(\epsilon\Delta)} &=&
  \,- \, \Frac{1}{2}   \epsilon  \mF_C'
\, \bra    u_\mu \nabla^\mu (\mF_C^{-\frac{1}{2}} \Delta) \ket
+ \mF_C^{-\frac{1}{2}} \epsilon \bra \Delta J_P'\ket
\, ,
\eear
in terms of $\mK= \mF_C^{-1/2} \mF_C'$ and $\Omega=2  \mF_C''/\mF_C - (\mF_C'/\mF_C)^2$.
Through a proper definition of the
differential operator $d_\mu \vec{\eta}=\partial_\mu \vec{\eta} + Y_\mu
\vec{\eta}$, one can rewrite $\mL_2^{\cO(\eta^2)}$ in the canonical form
\bear
\mL_{2}^{\cO(\eta^2)}   &=& - \Frac{1}{2} \vec{\eta}^T\, (d_\mu d^\mu +\Lambda) \vec{\eta}
\label{eq.Lagr-quad-fluctuation}
\eear
where $d_\mu$ and $\Lambda$ depend on ${h}$, ${u}_{L,R}$ and on the gauge boson and fermion fields.

The  $\mL_2^{\cO(\eta^2)}$ term in the generating functional is just a
Gaussian integral over the
quantum fluctuations $\vec{\eta}$, which can be performed with standard methods
to provide the 1-loop effective action~\cite{book-Donoghue,heat-kernel}:
\bear
\label{eq.div}
S^{1\ell}    
&=& \Frac{i}{2}    {\rm tr}      \, \log{ \left(d_\mu d^\mu +\Lambda\right)}  \, ,
\eear
where tr  stands for the full trace of the operator, also in coordinate space.
One can then extract the residue of the $1/(d-4)$ pole in dimensional
regularization using the Schwinger--DeWitt proper-time representation of
the operator in Eq.~(\ref{eq.div}) and the heat-kernel
expansion~\cite{heat-kernel}:
\bear
S^{1\ell}    
&=& \,-\, \divergence \, \Int {\rm d^d x} \,\,\,   {\rm Tr}
\left\{ \Frac{1}{12} Y_{\mu\nu} Y^{\mu\nu} +\Frac{1}{2}\Lambda^2
\right\}  \, +\, {\rm finite}
\nn\\
&=&
\,-\, \divergence \, \Int {\rm d^d x} \,\,\, \sum_k \, \Gamma_k \, \mO_k \,
+\, {\rm finite}\, ,
\label{eq.1loop-div}
\eear
with $\divergence =  [16\pi^2 (d-4)]^{-1} \mu^{d-4}$.
The divergence is determined by the non-derivative quadratic fluctuation $\Lambda$
and the differential operator $d_\mu$ through
$Y_{\mu\nu} =[d_\mu, d_\nu]=\partial_\mu Y_\nu -\partial_\nu Y_\mu
+[Y_\mu,Y_\nu]$,
and we note that both $\Lambda$ and $Y_{\mu\nu}$ are $\cO(p^2)$ in the
chiral counting.
In Eq.~(\ref{eq.1loop-div})  Tr  refers to the trace over the $4\times 4$
operators that act on
the fluctuation vector $\vec{\eta}^T=(\Delta_i,\epsilon)$.
The basis of local operators $\mO_k$ that covers the space of one-loop
divergences contains purely bosonic terms (given in Table~\ref{tab.div}) and
operators including
fermions (discussed later in Eq.~(\ref{eq.fermion-div})).
For the UV-divergent  part of the effective action
we have a chiral expansion in powers of $p$ similar to that for the EFT
Lagrangian in (\ref{eq.EFT-Lagr}): $\mL^{1\ell ,\, \infty} =
\mL_2^{1\ell ,\, \infty}  \,+\,
\mL_4^{1\ell ,\, \infty} \, +\, ...$

The UV divergences with the structure of the $\mL_2$ operators
in Eq.~(\ref{eq.L2}) are given by
\bear
\mL_2^{1\ell,\, \infty} &=&
  - \divergence
\bigg\{
\Frac{1}{8}\left[ \Frac{ \mF_C' V'}{\mF_C}(4-\mK^2)
- \mF_C\Omega V''\right] \bra u_\mu u^\mu\ket
\nn\\
&&  -\, \frac{3\mF_C' V' \Omega}{8\mF_C}
\left(\Frac{\partial_\mu h}{v}\right)^2
 +\left[\Frac{1}{2}\left( V''\right)^2
+ \Frac{3 \mK^2}{8  \mF_C    }  \left(V' \right)^2 \right]
\nn\\
&&
+ \left( V'' \bra  J_{S}''     \ket
 - \Frac{3 \mF_C' V' }{2\mF_C} \bra \Gamma_{S} \ket  \right)\bigg\}
\, ,
\label{eq.L2-div}
\eear
where we define
$\Gamma_{S}= \mF_C^{-1} ( J_{S} - \mF_C' J_{S}' /2)$.
These UV divergences are cancelled out through the renormalization of the
various parts
of $\mL_2$:
the couplings in the $\mF_C$ term (1st line);
the Higgs kinetic term (1st term in 2nd line),
which requires a NLO Higgs field redefinition;
the coefficients of the Higgs potential, e.g. the
Higgs mass (2nd bracket in 2nd line); the Yukawa term couplings in $Y$ (3rd line).

The $\cO(p^4)$ divergences $\mL_4^{1\ell,\infty}$
are split here into two types, according to whether they include fermion
fields or not. The purely bosonic
$\cO(p^4)$ divergences $\mL_4^{1\ell,\infty}|_{\rm bos}$ are
summarized in table~\ref{tab.div},
where we provide the factors $\Gamma_k=\Gamma_k[h/v]$
corresponding to each $\cO(p^4)$
operator $\mO_k$ in the effective action.

The structure of $\cO(p^4)$ UV divergences with fermion field operators
is slightly more involved:
\bear
&& \mL_4^{\rm 1\ell,\, \infty}|_{\rm ferm} =
  - \divergence
\bigg\{
\bra   \left(\Frac{\mK^2}{4}-1\right)  \Gamma_{S}
 - \Frac{\mF_C \Omega }{8}  J_{S}''\ket \, \bra u^\mu u_\mu \ket
\nn\\
&& + \Frac{3}{4} \Omega  \bra \Gamma_{S}\ket \, \left(\Frac{\partial_\mu
h}{v}\right)^2
+ \Frac{1}{2} \Omega  \bra \Gamma_{P} u^\mu \ket \,
\left(\Frac{\partial_\mu h}{v}\right)
\nn
\label{eq.fermion-div}
\\
&&
 + \Frac{1}{2}  \bra {J_{S}''}\ket^2
+ \Frac{3}{2}   \bra \Gamma_S\ket^2
+ \frac1{\mF_C} \left( 2  \bra\Gamma_{P}^{\,2} \ket
-    \bra \Gamma_{P}\ket^2\right)
\,\bigg\} .
\eear
with  $\Gamma_P=  J_{P}' - \mF_C^{-1} \mF_C' J_{P} /2$.

Any operator not listed in Eqs.~(\ref{eq.L2-div})
and~(\ref{eq.fermion-div}) and Table~\ref{tab.div}
is not renormalized at one-loop by scalar boson loops.
In the SM limit one finds $\Omega=0$, $\mK=2$ and $J_{S}''=\Gamma_{S,P}=0$,
so all the $\mL_4^{1\ell,\,\infty}$ operators in Eq.~(\ref{eq.fermion-div}) and
Table~\ref{tab.div}
vanish but for  $\Gamma_{10}^{\rm SM}=-1/6$, 
which turns out to be independent of the Higgs field and
is absorbed through the renormalization of $g$ and $g'$ in $\mL_{YM}$.
Furthermore, apart of $\mL_{YM}$, only the non-derivative operators (the Yukawa
term and Higgs potential)
get renormalized due to the scalar loops in the SM limit.

\section{Renormalization}

In order to have a finite 1-loop effective action
the divergences in Eq.~(\ref{eq.1loop-div}) are canceled by
the counterterms
\bear
\mL^{\rm ct}  
&=&  \sum_k \, c_k \, \mO_k\, ,
\eear
with the renormalization condition
$c_k = c_k^r + \divergence \, \Gamma_k$ .

The $\Gamma_k$'s have a Taylor expansion of the form
$\Gamma_k[h/v]=\sum_n \Gamma_{k,n} (h/v)^n/n!$,
and similarly, $c_k[h/v]=\sum_n c_{k,n} (h/v)^n/n!$.
The couplings of ${\cal O}(p^4)$ operators not present in
Eqs.~(\ref{eq.L2-div}) and~(\ref{eq.fermion-div})
do not get renormalized by scalar loops.
This leads to the renormalization group equations
for the $\cO(p^4)$ coefficients,
\bear
\Frac{\partial c_{k,n}^r}{\partial\ln\mu } &=&
- \, \Frac{\Gamma_{k,n}}{16\pi^2}\, .
\eear
Physically, this means that the NLO effective couplings
will appear in the amplitudes in combinations with logarithms of energy scales $E$
in the form
\bear
\mM_{\cO(p^4)} &\propto &  \left(  c_{k,n}^r(\mu) - \Frac{\Gamma_{k,n}}{16\pi^2}
 \ln\Frac{E}{\mu}
\right)\, E^4\, .
\eear
As it is well known from  Chiral Perturbation Theory,  the size
of these logs is not
known a priori
and can even be more dominant than the ${\cal O}(p^4)$ finite pieces.
For instance, in QCD for $\mu\sim M_\rho$,
the logarithms in the pion vector form-factor are numerically subdominant in
comparison
with the tree-level contributions, whereas one finds the opposite
behavior in $\pi\pi$
production
in the scalar-isoscalar channel.
There is a priori no reason to neglect the running of the coefficients
in the ECLh Lagrangian when confronting the experimental data
against the (nonlinear) effective description of EWSB.

In Table~\ref{tab.div} we have given the $\Gamma_k$ that provide
the full renormalization of the purely bosonic $\mL_4$ Lagrangian
for an arbitrary number of Higgs fields. In the last
column one can find their contributions with the minimal number of Higgs fields,
$\Gamma_{k,0}=\Gamma_k[0]$, which we have used as a check of our results.
In the $\mF_C=1$ limit ($a=b=0$) we recover the running
of the Higgsless EW chiral Lagrangian~\cite{Morales:94,Longhitano:1980iz}.
Part of our results for the 1-loop running have been already determined in the general
ECLh case in $WW,\, ZZ,\, hh$-scattering
($c_{4,0}$,..., $c_{8,0}$)~\cite{1loop-WW-scat}
and $\gamma\gamma$-scattering and related photon processes ($c_{1,0}$,
$c_{2,0}-c_{3,0}$)~\cite{1loop-AA-scat}.
Although the corresponding experimental analyses are limited so far by
statistics and yield
very loose constraints on these couplings~\cite{LHC-exp,WW-exp,AA-exp}, their
accurate determination or the feasibility to set more stringent bounds
in the future
requires a careful control of these $\cO(p^4)$ loop corrections.

Since the operators proportional to $A_{\mu\nu} A^{\mu\nu}$ and $A_{\mu\nu} Z^{\mu\nu}$
are only contained in the combination
$\Delta\mL=    (c_1/4 +c_{10}/2)    \bra f_+^{\mu\nu} f_{+\, \mu\nu}\ket\subset \mL_4$,
the NLO couplings that describe the vertices
$h\to\gamma\gamma$ and $h\to \gamma Z$
are renormalization group invariant, as was already found
in ~\cite{1loop-AA-scat} and~\cite{Azatov:2013}, respectively.
As $\Gamma_{1}/4 +\Gamma_{{10}}/2 =-1/12$ is independent of $h$,
a similar thing applies to $\gamma \gamma$ and $\gamma Z$ vertices with more Higgs fields
($\gamma\gamma,\, \gamma Z \to hh, \, hhh\, ...$).

Deviations from the SM at LO (e.g., by having a value $a\neq 1$  in $\mF_C$,
that modifies the $hWW$ coupling)
would also imply the appearance of the four-fermion UV divergences
in Eq.~(\ref{eq.fermion-div}), and thus the contribution of the associated
chiral logarithms to flavor-changing neutral current processes.
In addition to the usual bounds on the corresponding four-fermion couplings,
the study of these transitions may set strong constraints on the LO couplings in nonlinear Higgs EFT.

Let us finally observe that the Higgs potential gets divergent
corrections proportional to $(V')^2$ and $(V'')^2$, {\it i.e.} proportional to $m_h^4$,
that could be relevant in the study of the stability of the Higgs potential,
a subject beyond the scope of this letter.

This article has been focused on the 1-loop contributions of
SM scalar particles and the induced renormalization at NLO in the chiral counting.
Because scalars couple derivatively in nonlinear EW models,
the scalar loops are the only source of NLO divergences
that scale like the fourth power of the external momenta,
$(q_i)^4$, e.g. $c_8$~in~Table~\ref{tab.div}.        
The fermionic operators in    $\mL_4^{1\ell,\infty}|_{\rm ferm}$
in~(\ref{eq.fermion-div})
scale as $(q_i)^3$ and $(q_i)^2$ (recall that the fermion field scales as the square root
of the external momenta~\cite{Hirn:2004,EW-chiral-counting,Georgi-Manohar}),
with the remaining powers of $p$ given by the fermion masses in our computation.
An analogous thing happens with the $\mL_2$ renormalization in
Eq.~(\ref{eq.L2-div}):
the operators therein formally scale like $\cO(p^4)$ but,
since they are proportional to $V'$ or $V''$, at least two
of the powers of $p$ are actually $m_h^2$.
On the other hand,
contributions from  gauge bosons and fermions inside the loop,
which are not included in this work, will produce
UV divergences of order $(q_i)^3$ and $(q_i)^2$, as these particles
couple non-derivatively and proportionally
to the $g$ and $g'$ gauge couplings and Yukawas $y_{t,b}$
({\it i.e.}, proportionally to the gauge boson and fermion masses.)
The computation and analysis of the latter is postponed to future work.

{\bf Acknowledgements:}\\
We would like to thank J.F.~Donoghue,
M.J. Herrero and A. Pich for discussions on the heat-kernel expansion
and EW chiral Lagrangians,
and S.~Saa for pointing out the simplification in the structure of the Higgs kinetic term.
PRF thanks P. Ruiz-Torres for stimulating and cheerful discussions.
The work of FKG is supported in part by the NSFC and DFG through funds
provided to the Sino-German
CRC 110 ``Symmetries and the Emergence of Structure in QCD"(NSFC Grant No.
11261130311) and
NSFC (Grant No. 11165005),       
the work of JJSC is supported by ERDF funds from the European Commission
[FPA2010-17747, FPA2013-44773-P, SEV-2012-0249, CSD2007-00042],   
and the research of PRF was supported by the Munich Institute for Astro- and Particle Physics (MIAPP)
of the DFG cluster of excellence "Origin and Structure of the Universe.    


\end{document}